\documentstyle[twocolumn,epsfig]{article} 
\begin{document}
 
\title{Tests of a High Resolution Time of Flight System \\
         Based on Long and Narrow Scintillator}

\author{E.~Chen, M.~Saulnier, W.~Sun and H.~Yamamoto.\\
       Harvard University, Cambridge, Massachusetts 02138}

\twocolumn[\maketitle
\parbox{\textwidth}{We have tested a prototype time-of-flight 
system based on bulk scintillator block of dimensions 
$2.5 \times 2.5 \times 200$ cm$^3$. Using a calibration scheme similar to the
one used in actual collider experiments, we have achieved a resolution
of 71 ps using Amperex XP2020/UR photomultipliers
and 81 ps using proximity-focusing fine-mesh photomultipliers 
(Hamamatsu R2021).  
Results are also obtained for 
scintillating fiber blocks of the same dimensions.
Good internal reflectivity of the bulk scintillator block resulted in
resolutions superior to the fibre blocks.
A single-photon pulsed laser system was used to study photomultipliers
and the results were used in a Monte Carlo simulation of the
system to study the critical elements that determine the resolution.}
\vskip0.5in]

\section{Introduction}

In many high energy
physics experiments the identification of particle species 
is accomplished through time-of-flight (TOF) measurements utilizing a 
system of plastic scintillators read out by photomultiplier tubes (PMT's).
A typical configuration
for a detector at a colliding beam facilities is a barrel geometry 
consisting of a single layer
of long rectangular scintillation counters parallel to the beam, segmented 
in azimuth, and read out
at both ends by photomultipliers coupled to light guides~\cite{typical}.  
Two important factors that define the performance of such TOF 
systems are the timing resolution and the
efficiency to register a clean hit. Using 2.8 meter long 
counters of thickness 5 cm and width 10 cm, the CLEO II experiment obtains 
resolutions  
of 139 ps for bhabha events and 154 ps for hadronic events
\cite{cleoiinim,Giles}.
The efficiency for obtaining good timing information for a track is limited
primarily by events where two 
or more particles strike the same counter.  In 
these cases one is no longer able to effectively utilize 
the timing measurements from 
both ends, and the resolution is consequently substantially worse.  
The efficiency at 
CLEO II for spherical events is approximately 85\% and for jetlike events, 
approximately 75\%. 

The factors that determine the timing resolution are quite well understood:
the number of photoelectrons detected,~\cite{Atwood} 
the dispersion of path length from the scintillation point to the 
photocathode,
the scintillation decay time, 
and the transit time spread of the photomultiplier. 
Thus, one tries to use fast 
and efficient scintillator with long attenuation length,
fast photubes with small transit time spread covering as much area
as possible at the end of the scintillator preferably without any
light guides. 
As a rule of thumb, one loses the precious scintillation photons
by a fraction proportional to the ratio of the cross section of the
scintillator to the photocathode area. The light guide can add dispersion
of photon path length and further photon loss~\cite{lightguides}.
The lack of light guides usually requires
photomultipliers that can operate within a high magnetic field ($\sim$1 T). 
The thickness of the scintillator should be as thick as
possible as long as other detector requirements are met~\cite{thick}.

The efficiency argues for a fine segmentation. The effective attenuation of
a finely segmented system, however, tends to be small due to the
increased number of internal reflections.
It has been suggested that resolutions for long counters can be improved
by using scintillating fibers~\cite{Kuhlen}. The advantage of a fiber counter 
is that the resolution at large distances is not 
severely limited by photon path length dispersion
and that the loss due to internal reflections are small~\cite{coreclad}.  
On the other hand, fiber counters have a small trapping 
angle which limits the absolute amount of light collected by the PMT.  
The authors of reference~\cite{Kuhlen} compared bulk and fiber 
counters and concluded that the system using the fiber counter has a better 
resolution at long distances.  
However, if the effective 
attenuation length of the bulk counter is made comparable to 
that of the fiber counter, then the bulk counter would give the superior 
resolution even at long lengths. 

In this article, 
we report a study of a prototype time of flight system using long and
narrow scintillator counters.  
Such counters could be arranged in a
two layer system where the inefficiency due to overlapping tracks would be 
substantially reduced.
Moreover, the timing resolution of such double-layer system  
would be roughly factor of $\sqrt2$ better than that of a single-layer
system\cite{othereffects}. In such system, one also need to deal with
the technique to install PMT's whose diameters are larger than the
transverse size of the scintillator block, and to 
combine multiple measurements for a given track. These are sujects of
future studies.

\section{Experimental setup}

The three different scintillator counters studied have the same dimensions of 
2 meters $\times$ 2.5 cm $\times$ 2.5 cm.  The first was 
bulk Bicron BC-408 scintillator (decay time 2.1 ns) 
wrapped in aluminized mylar.  The second was
a 6 $\times$ 6 bundle of 4 mm diameter Bicron BCF-10 scintillating fibers 
(decay time 2.4 ns) with a polystyrene-based core and 
polymethylmethylacrylate cladding, glued together with epoxy.     
The indices of refraction were 1.60 for the core and 1.49 for the clad.  The 
third counter was similar to the second, except that it was composed of
2 mm diameter fibers.  For both types of fiber, the clad thickness 
was 3\% of the core diameter.
The bulk scintillator 
was initially cast between glass planes.  The unfinished 
surfaces were then milled by diamond fly cutting~\cite{diamondmill}.  
This finishing technique yields a reflectivity superior to that obtained by 
standard polishing, achieving an improved effective attenuation length 
of the counter. A study of reflectivity of NE110 
scintillator~\cite{Kittenring} shows that any polishing dramatically
deteriorates the reflectivity. The ideal fabrication method would be to
cast all four lengthwise surfaces against glass plates.

\begin{figure}
 \centering
 \mbox{\psfig{figure=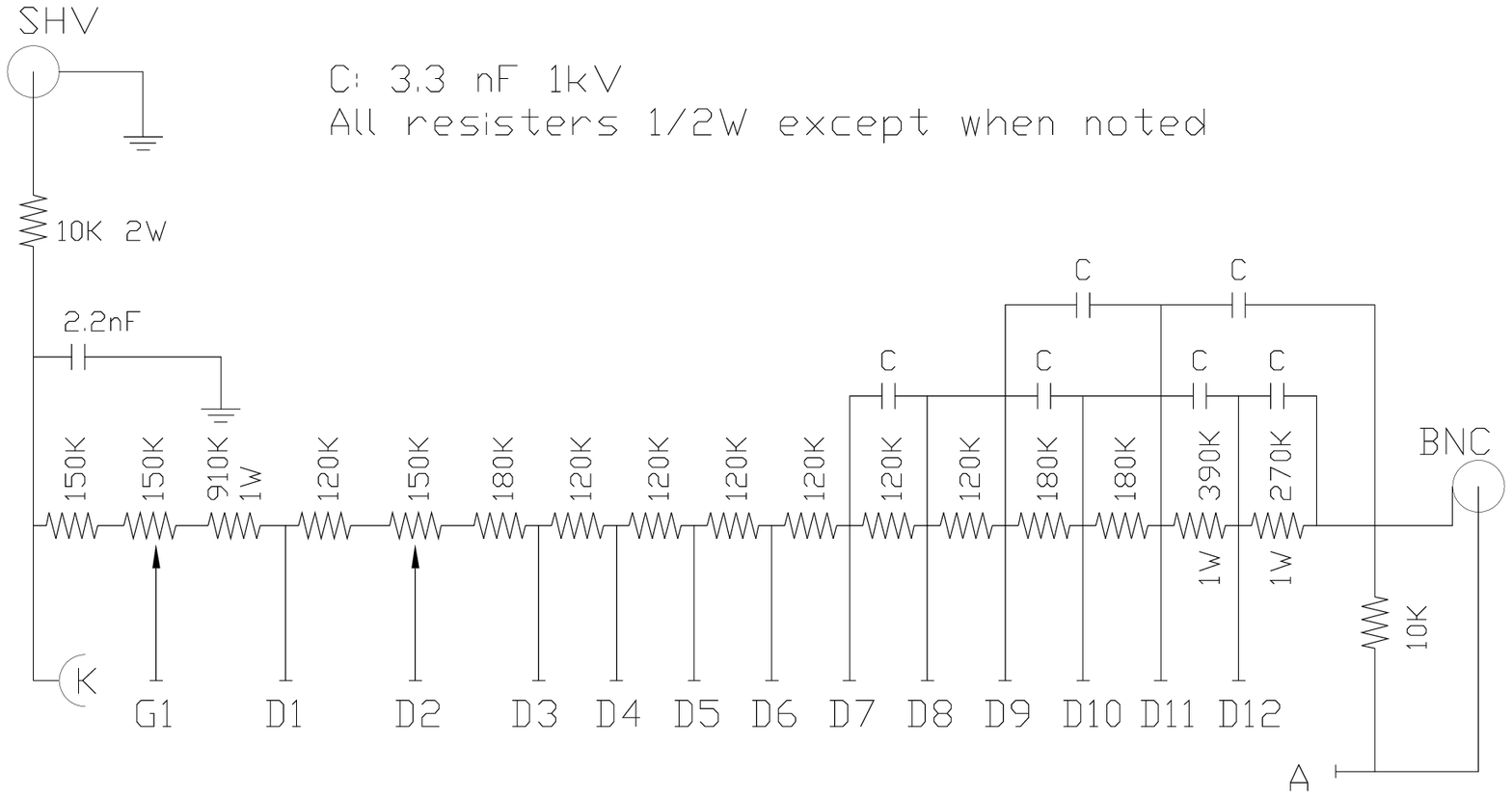,height=1.8in,width=3.2in}}
 \caption{Base design for the Amperex XP2020/UR.}
 \label{fig:corel_base2}
\end{figure}

Two different types of PMT's were used to detect the 
scintillation light from the test counter.  The first was a 
modification of the popular Amperex XP2020 called the XP2020/UR.  It has
a 12 stage dynode and a bialkali photocathode of useful diameter 4.5 cm.  
The XP2020/UR differs from the XP2020 in the internal connections of 
accelerating and focusing grids and it allows a higher
voltage to be applied between the cathode and the first dynode 
to reduce the transit time.
The base design is shown in Figure~\ref{fig:corel_base2} 
\cite{besttiming}.  
A further improvement in resolution was obtained by 
adjusting the two potentiometers at G1 and D2 to maximize the potential
drop from K to G1 and from D1 to D2, respectively.  
As will be shown later in the single photoelectron study, the G1 voltage is
critical in making the transit time uniform across the photocathode
surface. We applied
$-3000$V to the XP2020/UR when it was used with a fiber counter, 
and $-2500$V with the bulk counter to avoid saturation.  Such a tube
would have to be placed outside the magnetic field of a detector, and would
be coupled to the counters via light guides.  

\begin{figure}
 \centering
 \mbox{\psfig{figure=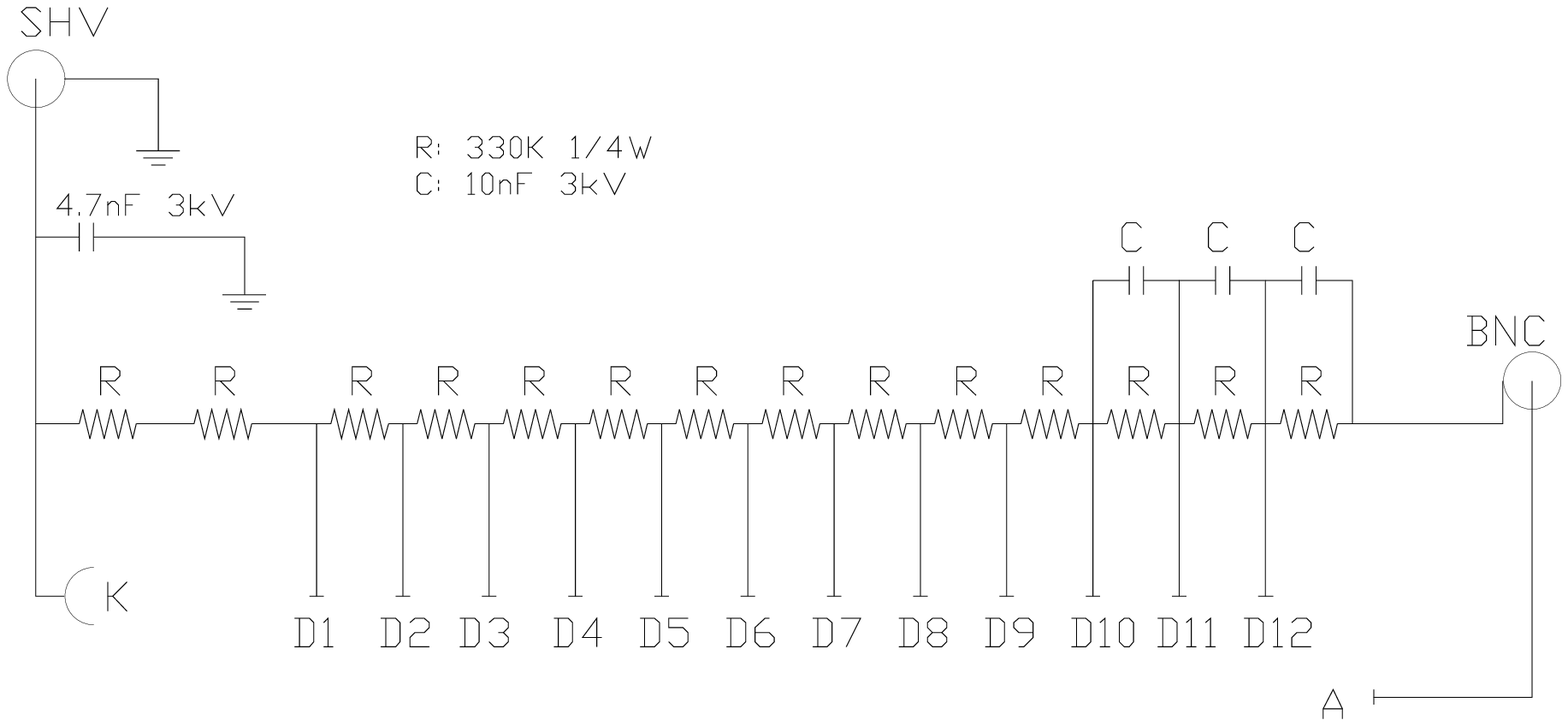,height=1.6in,width=3.2in}}
 \caption{Base design for the fine-mesh photo multiplier Hamamatsu R2021.}
 \label{fig:corel_base1}
\end{figure}

The second type was a 
Hamamatsu R2021 proximity-focusing fine-mesh PMT which can be
placed inside the detector's magnetic field.  This PMT has a 12 
stage dynode structure and a 2.2 cm diameter bialkali photocathode.  
The gain is specified to be $5 \times 10^4$ at 0 gauss, and 
$1.5 \times 10^3$ in a 10 kilogauss field parallel to the axis 
of the PMT.  We amplified the outputs of the R2021's with Mini-Circuits
ZFL1000-LN low noise $\times 10$ amplifiers.  The base design for this 
PMT is shown in Figure~\ref{fig:corel_base1}.
When a fine-mesh PMT is operated inside a high magnetic field, the primary
effects of the field are (a) to increase the probability of electrons 
to hit the wire of the
mesh due to the spiraling motion along the field, (b) 
reduce the probability of secondary
emission electrons to reach the next stage, and that (c) the cloud of
amplified electrons follows roughly the direction of magnetic field as it
passes through the layers of meshes~\cite{meshstudy}.
The effect (b) is the main reason for gain loss in a high field, while
the effect (a) result in higher gains when the PMT axis is 
at some angle $\theta$ (about 30 degrees)
with respect to the field than zero angle. The effect (c) is responsible
for the sharp drop of gain for $\theta$ greater than some value (about 50
degrees).
We have not studied fine-mesh PMT's inside high magnetic field.
Studies show, however, that the deterioration in timing resolution
is not large~\cite{bfield,STAR}.

\begin{figure*}
 \centering
 \mbox{\psfig{figure=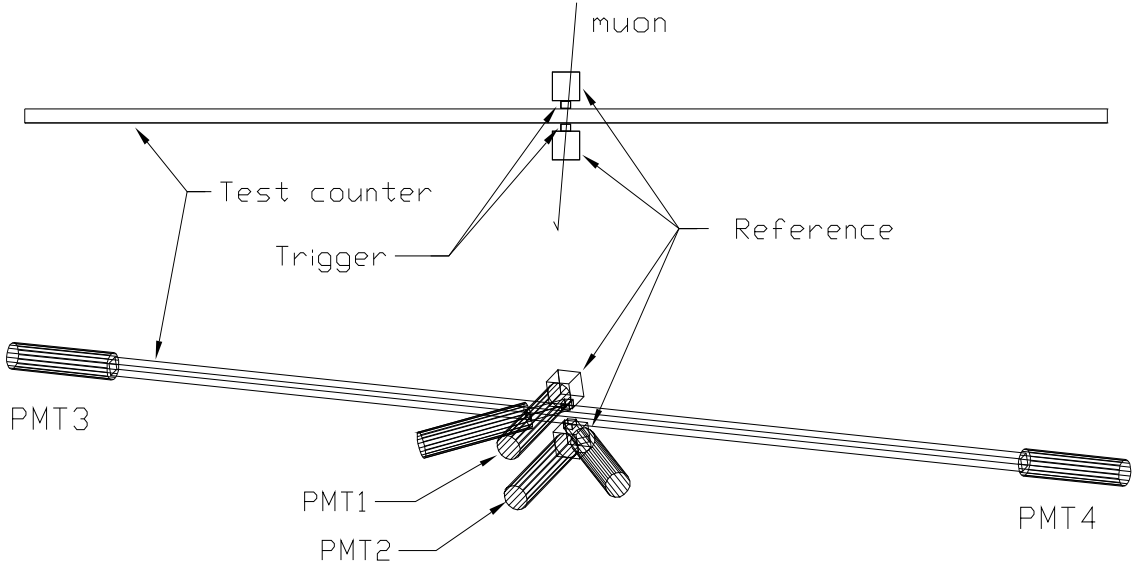,height=3.0in,width=6.5in}}
 \caption{Cosmic ray telescope test setup.}
 \label{fig:TOF-CRT}
\end{figure*}

The scintillator and PMT to be tested were placed in a cosmic ray telescope 
as shown in Figure~\ref{fig:TOF-CRT}, 
optically coupled using an UV-transparent grease GE Viscasil 60000. 
The cosmic ray events were selected 
by coincidence of the trigger counters with dimensions $2\times2\times1.2$
cm$^3$ each.  The reference counters positioned 
next to the trigger counters provided a measurement of the time that the
cosmic ray passed through the test counter.  All four reference
and trigger counters were mounted on a trolley which could be positioned 
anywhere along the test counter.

The coincidence of trigger PMT pulses 
defined the TDC start time and generated a
gate for the ADC.  The ADC pedestal was measured every five minutes.  
Each pulse from the two reference PMT's and the two test PMT's was split 
and passed to both a LeCroy 2249A ADC and leading edge discriminators
(LeCroy 623B and 4608C) 
whose output was sent to a LeCroy 2228A TDC calibrated to 50 ps/count.  
The discriminator threshold for the reference counters was chosen to be 
$-30$ mV, which is roughly 5\%\ of the typical pulse height,
in order to maximize acceptance.  
The threshold for the trigger 
counters was chosen to be $-400$ mV, which was roughly 1/3 of 
the typical pulse height in order to exclude background events.

\section{Attenuation length}

\begin{figure}
 \centering
 \mbox{\psfig{figure=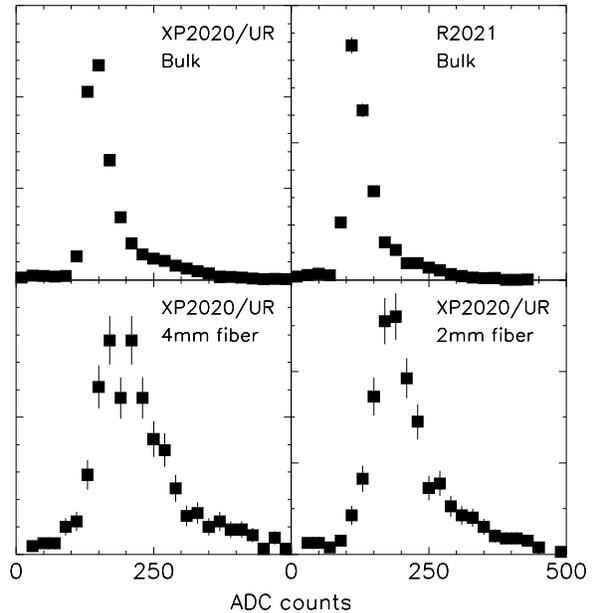,height=3.2in,width=3.2in}}
 \caption{Distribution of ADC counts for a test counter photo multiplier
(PMT3) for four different configurations:
the XP2020/UR's on the bulk counter, the R2021s on the bulk 
counter, the XP2020/UR's on the 4mm fiber counter, 
and the XP2020/UR's on the 2mm fiber counter.}
 \label{fig:091995_01}
\end{figure}

Figure~\ref{fig:091995_01} shows the pulse height distributions of the 
test counter PMT's. The trolley 
was positioned at the center.  
Results are shown for 
four different combinations of PMT's and counters: 
the XP2020/UR's at $-2500$V on the bulk counter, the R2021s on the bulk 
counter, the XP2020/UR's at $-3000$V on the 4 mm diameter fiber counter, 
and the XP2020/UR's at $-3000$V on the 2 mm fiber counter. 

\begin{figure}[t]
 \centering
 \mbox{\psfig{figure=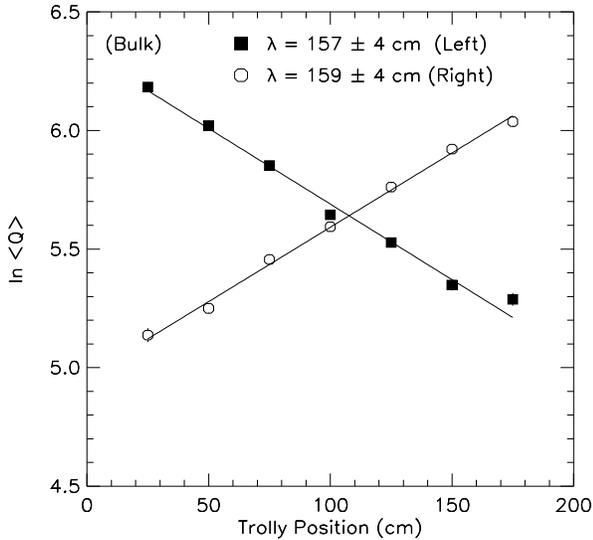,height=2.8in,width=3.2in}}
 \caption{Attenuation length of the bulk counter, determined by 
fitting the mean pulse height from each end to an exponential.}
 \label{fig:073195_04}
\end{figure}

The effective attenuation length of each counter was measured by
fitting a exponential function to  
the pulse height as a function of the distance from the each end of the
counter. This gives two independent measurements for a single scan.  
Figure~\ref{fig:073195_04} shows the results obtained for the bulk
counter, indicating an effective attenuation length of 160 cm.  
For the fiber counters, we find an effective attenuation length 168 
cm for the 4 mm diameter fiber counter, and 147 cm for the 2 mm diameter
fiber counter. 
Note that the effective attenuation length of the bulk counter is 
comparable to the effective attenuation lengths of the fiber counters.  
Thus, relative to the fiber counters, 
the bulk counter does not suffer from the loss of
photon statistics at the locations far from PMT. As a result, the bulk
counter may have an overall timing resolution superior to those of the
fiber counters due to the better photon statistics, which we observe to
be the case as reported below.
 
\section{Timing resolution}

In the following discussion, we assign the numbers 1 and 2 to the reference 
counter PMT's, 3 to the test counter PMT at $z = 0$ cm, and 4 to the test 
counter PMT at $z = 100$ cm. 
The mean time measured by the two reference counters is the time at which the
cosmic ray crossed the test counter.  The time measured by PMT $i$ 
must be corrected to account for the time required for the pulse to reach 
the discriminator threshold (the 'time-walk' correction). 
The correction has the form
$$ T_i' = T_i - a_i Q_i^{-b_i}$$
where $T_i$ is the measured TDC time and $Q_i$ is the pedestal-subtracted ADC
count.  We use $b_1 = b_2 = 0.5$ for the reference PMT's.  The constants 
$a_1$ and $a_2$ are determined from the data by minimizing the width of 
the distribution 
$$T_{{\rm ref}-} \equiv (T_1' - T_2')/2.$$  
We collect data with the trolley placed at 25 cm intervals along the test
counter, from $z$ = 25 cm to $z$ = 175 cm.  We combine the data from all
$z$ positions and determine a single value for $a_1$ and for $a_2$ which is
independent of $z$.

\begin{figure}
 \centering
 \mbox{\psfig{figure=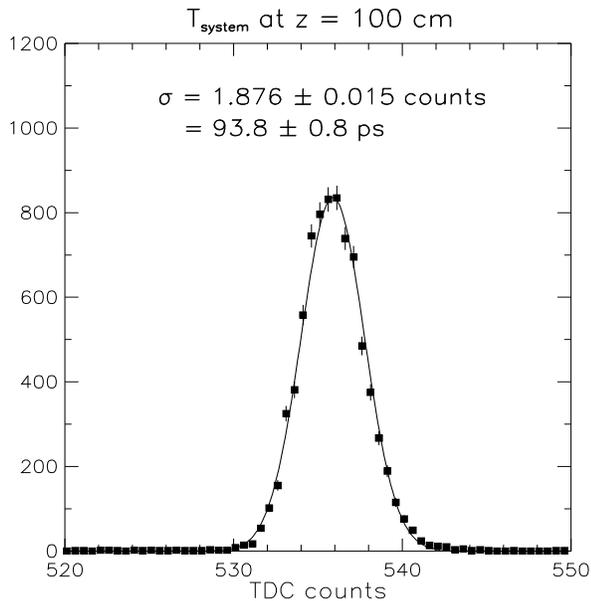,height=3.2in,width=3.2in}}
 \caption{Distribution of $T_{\rm system}$ for the trolley at 
$z = 100$ cm, with the XP2020/UR's on the bulk counter.  The reference
counter resolution must be subtracted in quadrature from the width of 
this distribution to obtain $\sigma_{\rm tof}(z=100)$.}
 \label{fig:072095_02}
\end{figure}

Using the values for $a_1$ and $a_2$, we solve 
for the constants $a_3, a_4, c_3$ and $c_4$ of the test counter PMT's
by minimizing the widths of 
the distributions
$$ T_{\rm L}(z) \equiv T_3' -  c_3 z - T_{{\rm ref}+}$$
and 
$$ T_{\rm R}(z) \equiv T_4' -  c_4 z - T_{{\rm ref}+}$$
with
$$T_{{\rm ref}+} \equiv (T_1' + T_2')/2.$$
The constant $c_i$ is the inverse of the effective speed of light in the test 
counter and z is the distance to the PMT in question. Note
in particular that the constants $a_3$ and $a_4$ are independent of $z$.
We thus follow the method of calibration identical to the one 
used at the CLEO II detector where the beam crossing time is replaced
by $T_{{\rm ref}+}$.
We do not require that $c_3 = c_4$ following the standard 
calibration technique of CLEO II.  
The value of the time-walk constant
were chosen to be $b3=b4=0.15$ as described later. 
We define the time resolutions of the left and right PMT's to be
$$\sigma_3(z) = \sqrt{\sigma^2_{\rm L}(z) - \sigma^2_{\rm ref}}$$
$$\sigma_4(z) = \sqrt{\sigma^2_{\rm R}(z) - \sigma^2_{\rm ref}}.$$
where
$\sigma_{\rm L}(z)$ and $\sigma_{\rm R}(z)$ are the measured widths of 
$T_{\rm L}(z)$ and $T_{\rm R}(z)$, respectively.
We then construct 
the weighted average of crossing time measurements as follows:
$$T_{\rm system}(z) \equiv
\frac{T_{\rm L}(z)\sigma^2_4(z) + T_{\rm R}(z)\sigma^2_3(z)}
     {\sigma^2_3(z) + \sigma^2_4(z)}.$$
The measured width of this distribution is denoted as 
$\sigma_{\rm system}(z)$.  
Figure~\ref{fig:072095_02} shows $T_{\rm system}$
for the bulk counter at $z = 100$ cm, using XP2020/UR's.  
The distribution is well-described by a gaussian, and has a resolution of 
$94 \pm 1$ ps.  

\begin{figure}
 \centering
 \mbox{\psfig{figure=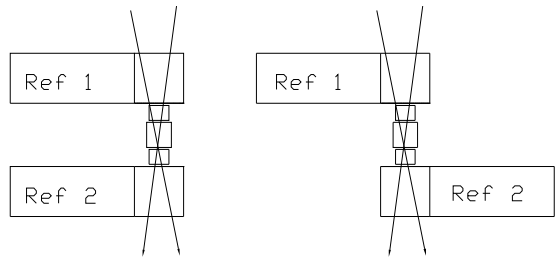,height=1.7in,width=3.2in}}
 \caption{Reference counter configurations. The standard configuration
 (left), and that for the measurement of reference counter resolution (right).
 The effect of timing correlation between Reference 1 and 2 due to the
 track path for T1 + T2 in the left configuration 
 is the same as that for T1 - T2 in the right.}
 \label{fig:ref-calib}
\end{figure}

\begin{figure}
 \centering
 \mbox{\psfig{figure=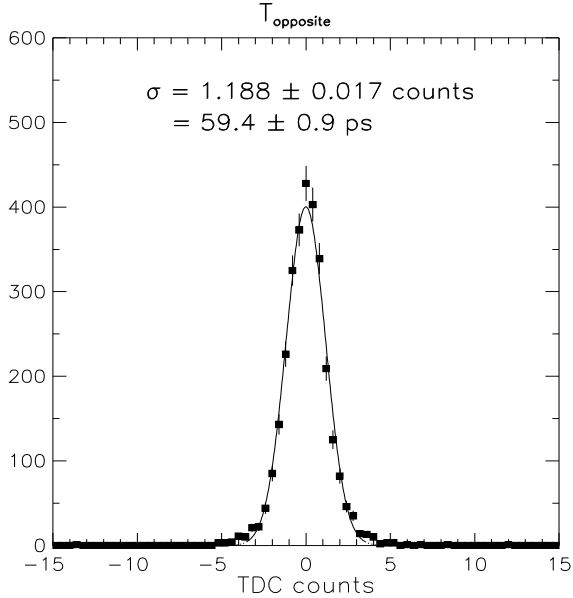,height=3.2in,width=3.2in}}
 \caption{Resolution of reference counters, defined as the width
  of ${T_{\rm ref}-} = (T_1' - T_2')/2$ when the reference counter
  PMT's are oriented in opposite directions.}
 \label{fig:071995_01}
\end{figure}

Finally, the contribution from the reference counters must be subtracted in
quadrature to obtain the resolution of the test counter:
$$\sigma_{\rm tof}(z) = \sqrt{\sigma^2_{\rm system}(z) - \sigma^2_{\rm
ref}}.$$
In previous studies of similar time-of-flight systems, 
the resolution of the reference counter time difference $T_{{\rm ref}-}$ was 
often taken to equal the 
average time $T_{{\rm ref}+}$.  However, such an 
assumption
can be made only if the cosmic ray tracks have a narrow angular deviation
and in the limit of small trigger counters.
With our geometry, the angular acceptance of cosmic rays is approximately
20 degrees from vertical. This gives rise to a correlation between
the two reference timings that cancels in $T_{{\rm ref}+}$ but not in
$T_{{\rm ref}-}$ since when the track is close to one of the PMT's then
it is away from the other (see Figure~\ref{fig:ref-calib}).
Under these conditions, the resolution of
$T_{{\rm ref}-}$ is significantly greater than the resolution of 
$T_{{\rm ref}+}$. There is also a smaller correlation introduced by
the variation of the track position in the mid plane between the two
trigger counters.
In order to properly account for these correlations, the reference
resolution was measured by placing
PMT's facing opposite directions.  
The width of 
$T_{{\rm ref}-}$ in this configuration should
be equal to that of $T_{{\rm ref}+}$ when both PMT's face the same 
direction. This assumption was verified by Monte Carlo 
studies~\cite{MCestimate}.  
The reference time
resolution was thus measured to be $\sigma_{\rm ref} = 59 \pm 1$ ps 
(Figure~\ref{fig:071995_01}).  
Using this value of $\sigma_{\rm ref}$,
we obtain the resolution of the test counter system to be 
$\sigma_{\rm tof} = 73$ ps at $z = 100$ cm.   The contribution to 
$\sigma_{\rm tof}$ from the finite TDC bin size is negligible.

\begin{figure}
 \centering
 \mbox{\psfig{figure=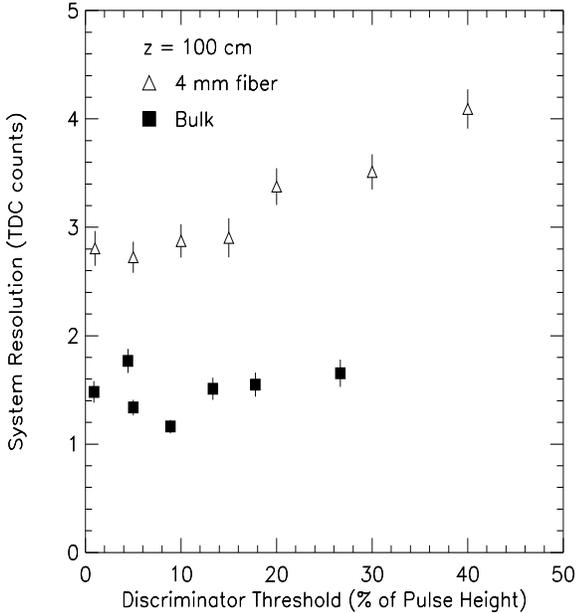,height=3.2in,width=3.2in}}
 \centering
 \caption{$\sigma_{\rm tof}(z = 100)$ as a function of discriminator 
threshold for test counter PMT's: bulk counter (solid square) 
and 4mm diameter 
fiber counter (open triangle).}
 \label{fig:073195_07}
\end{figure}

Having defined the timing resolution of the system, we examined the 
resolution obtained for different thresholds on the test counter 
discriminators.  For this study, data
were collected at $z = 100$ cm.  The test counter discriminator 
thresholds were set to a fixed percentage of the average pulseheight for 
each end, and runs were taken for various thresholds in the range
of 1\% to 40\% of the average pulse height. 
Figure~\ref{fig:073195_07} 
shows the resolution versus discriminator level for the bulk and 
the 4 mm fiber counters. 
The thresholds were chosen to be  
5\%\ for the 4mm fiber counter, and 10\%\ for the bulk counter.

\begin{figure}
 \centering
 \mbox{\psfig{figure=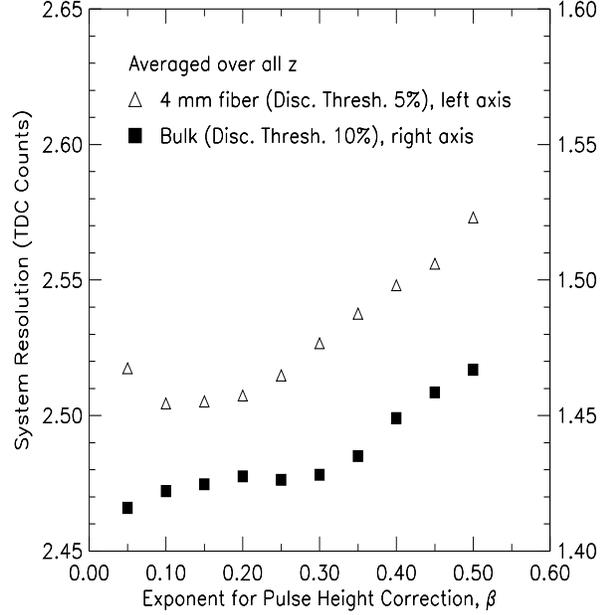,height=3.2in,width=3.2in}}
 \caption{$\sigma_{\rm tof}(z = 100)$ as a function of the exponential term
of the test counter time-walk correction $b_3 (= b_4)$: bulk counter 
(right axis) and 4mm diameter fiber counter (left axis).}
 \label{fig:073195_08}
\end{figure}

We next determined the value of the time-walk constants $b_3$ and $b_4$ 
constraining them to be equal.  For this study
data were collected for all $z$ positions, and the simple average of 
$\sigma_{\rm tof}(z)$ for all $z$ position was calculated.  
Figure~\ref{fig:073195_08}
shows how the average resolution depends on $b_3$ for the bulk and
for the 4 mm fiber counter, each at their optimal discriminator thresholds.
The data favor lower values than the oft-used value of 0.5.  
We choose $b_3 = b_4 = 0.15$
for both fiber and bulk counters using the XP2020/UR's.  

\begin{figure}
 \centering
 \mbox{\psfig{figure=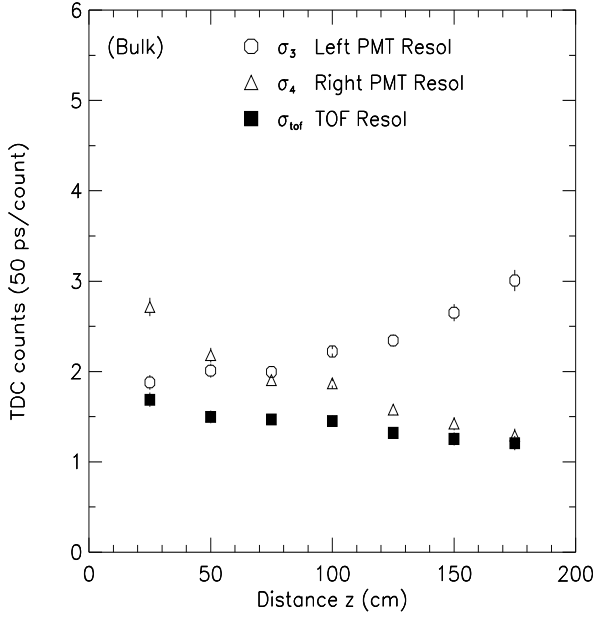,height=3.2in,width=3.2in}}
 \caption{Time resolutions of 
 the left PMT ($\sigma_3(z)$), the right PMT ($\sigma_4(z)$) and 
  the weighted average ($\sigma{\rm tof}(z)$). For the 
  bulk counter with XP2020/UR's.}
 \label{fig:073195_01}
\end{figure}

\begin{figure}
 \centering
 \mbox{\psfig{figure=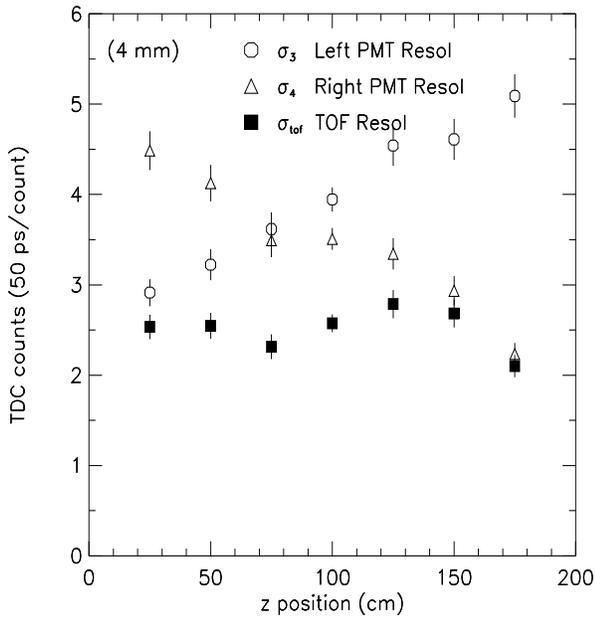,height=3.2in,width=3.2in}}
 \caption{Time resolutions of 
 the left PMT ($\sigma_3(z)$), the right PMT ($\sigma_4(z)$) and 
  the weighted average ($\sigma{\rm tof}(z)$). For the
4 mm diameter fiber counter with XP2020/UR's.}
 \label{fig:073195_02}
\end{figure}

\begin{figure}
 \centering
 \mbox{\psfig{figure=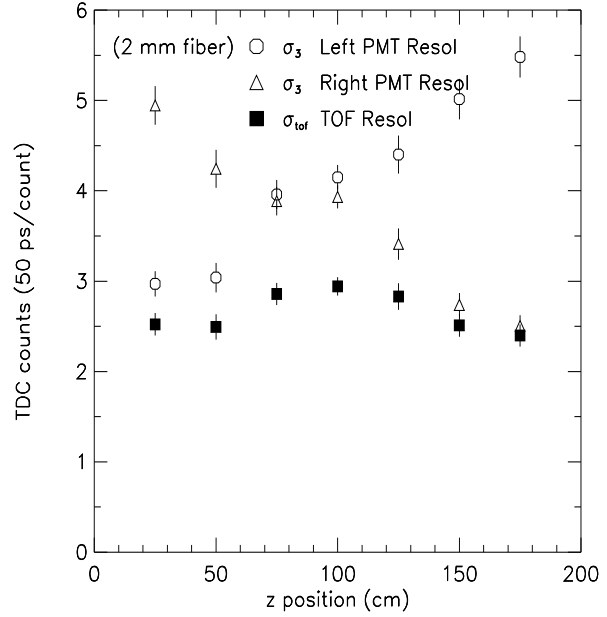,height=3.2in,width=3.2in}}
 \caption{Time resolutions of 
  the left PMT ($\sigma_3(z)$), the right PMT ($\sigma_4(z)$) and 
  the weighted average ($\sigma{\rm tof}(z)$). For the
  2 mm diameter fiber counter with XP2020/UR's.}
 \label{fig:073195_03}
\end{figure}

Figure~\ref{fig:073195_01} shows 
$\sigma_3(z)$, $\sigma_4(z)$, and $\sigma_{\rm tof}(z)$ for the bulk counter 
using the XP2020/UR's.
The left counter resolution was found to be worse than that of the right
counter. This was found to be consistent with the photon statistics as
measured and attributed to an imperfect optical coupling of the PMT to the
scintillator block.
We take the simple average of $\sigma_{\rm tof}(z)$ to obtain an overall
resolution of $71 \pm 1$ ps for the bulk counter. 
Figures~\ref{fig:073195_02} and \ref{fig:073195_03} show similar data 
for the XP2020/UR's on the 4 mm diameter and 2 mm diameter fiber 
counters, respectively.  
We find an average $\sigma_{\rm tof}$ of 
$125 \pm 3$ ps
for the 4 mm fiber counter and $133 \pm 2$ ps for the 
2 mm fiber counters.   

\begin{figure}
 \centering
 \mbox{\psfig{figure=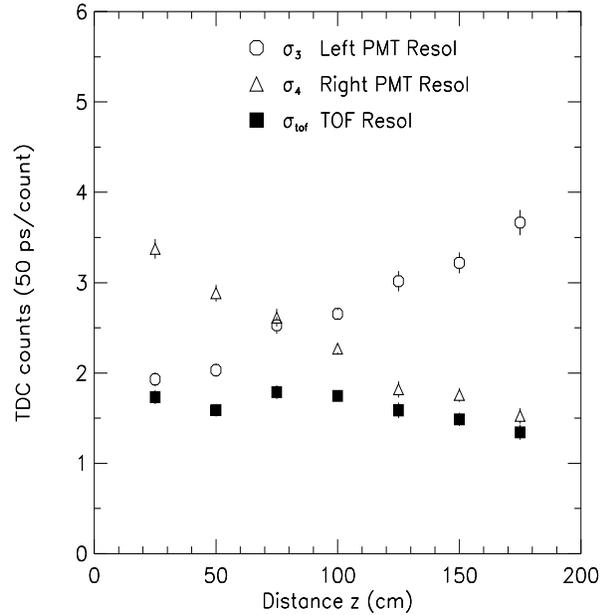,height=3.2in,width=3.2in}}
 \caption{Time resolutions of 
  the left PMT ($\sigma_3(z)$), the right PMT ($\sigma_4(z)$) and 
  the weighted average ($\sigma{\rm tof}(z)$).
  For the bulk counter with R2021s.}
 \label{fig:082395_01}
\end{figure}

Lastly, we studied the bulk counter using Hamamatsu R2021 proximity-focus
fine-mesh 
PMT's. For this study, we used the discriminator
threshold at 10\% of the pulse height. The optimal 
value for the constants $b_3$ and $b_4$ were found to be 0.30 using the method
described earlier.  Figure~\ref{fig:082395_01} shows 
$\sigma_3(z)$, $\sigma_4(z)$, and $\sigma_{\rm tof}(z)$
for the R2021s on the bulk 
counter.  
The average $\sigma_{\rm tof}$ is found to be $81 \pm 1$ ps which is
slightly worse than the resolution obtained by XP2020/UR. It should
be noted, however, that the photocathode of R2021 covered only 65\%\ of the
end of the bulk scintillator and the difference is consistent with the
photons lost. 
Table~\ref{tab:resols} summarizes the average $\sigma_{\rm tof}$ for each 
configuration.
\begin{table}
\begin{center}
\begin{tabular}{lr}
\hline
Configuration           & Resolution (ps) \\
\hline
XP2020/UR on bulk       & $71 \pm 1$  \\
XP2020/UR on 4 mm fiber & $125 \pm 3$ \\
XP2020/UR on 2 mm fiber & $133 \pm 2$ \\
R2021 on bulk           & $81 \pm 1$  \\
\hline
\end{tabular}
\caption{The measured time resolution $\sigma_{\rm tof}$ 
  for each configuration averaged over the length of the test counter.}
\end{center}
\label{tab:resols}
\end{table}

\section{Single photon study}

Examination of the single photon response of the XP2020/UR is instrumental
in understanding one of the components of the time-of-flight system resolution.
In particular, it is useful to know how the timing resolution varies according
to the position on the photocathode at which the photon is incident.  
In addition, it also allows 
the gain of the PMT/base combination to be determined
accurately. These are used as
input parameters for the Monte Carlo simulation of the system.

\begin{figure}
 \centering
 \mbox{\psfig{figure=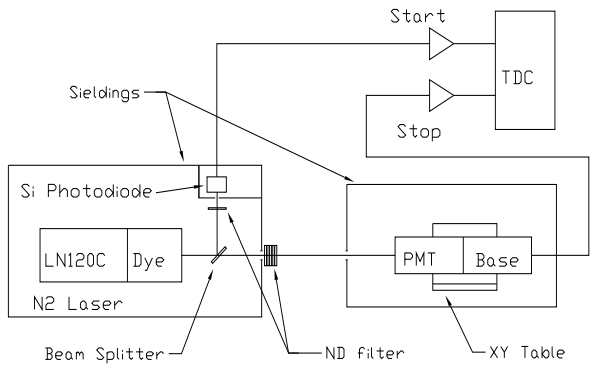,height=2.1in,width=3.2in}}
 \caption{The setup for the single photon study. The entire system except
     the TDC logic is placed in a dark room.}
 \label{fig:corel_sphoelec}
\end{figure}

The light source used for this investigation was a Laser Photonics LN120C
nitrogen laser, which delivered pulses of wavelength 337.1 nm and 
r.m.s. duration 70 ps, each with 70 $\mu$J, 
and at a repetition rate of about 10 Hz.  
After exiting the laser, the 
beam was split (see Figure~\ref{fig:corel_sphoelec}).  
Figure~\ref{fig:corel_sphoelec}.   
One beam triggered a fast photodiode 
(Newport 818 BB, with rise time $<$ 200 ps)
and defined the start 
time of the event while the other beam proceeded through a series of neutral 
density filters and then to the PMT.  The attenuation provided by the filters
was sufficiently large that a PMT pulse resulted only once for every 
$\sim$10 laser pulses.  Thus, approximately 
95\%\ of the signal events were due to one photoelectron.   

The PMT was mounted on a movable x-y table which allowed the position of the 
incident photon to be varied across the photocathode.  The PMT was 
magnetically shielded by mu metal inside degaussed soft iron.  An iris with a
1 mm diameter aperture was placed in front of the PMT to define 
the beam size at the photocathode.  
The signal from the PMT was sent to a Mini-Circuits ZFL-1000LN amplifier and 
then split to an ADC and TDC.
In order to shield the PMT from electrical noise generated by the laser, 
each of the two apparati was separately enclosed in an electrically grounded 
box of 1/8 inch-thick aluminum. 

The PMT was oriented such that longer dimension of the first dynode 
was horizontal. The repositioning of the stage was done in darkness 
to keep the PMT on and stable for the duration of the scan.  A common pulse 
height correction was applied to all runs from the scans of both axes.

\begin{figure}
 \centering
 \mbox{\psfig{figure=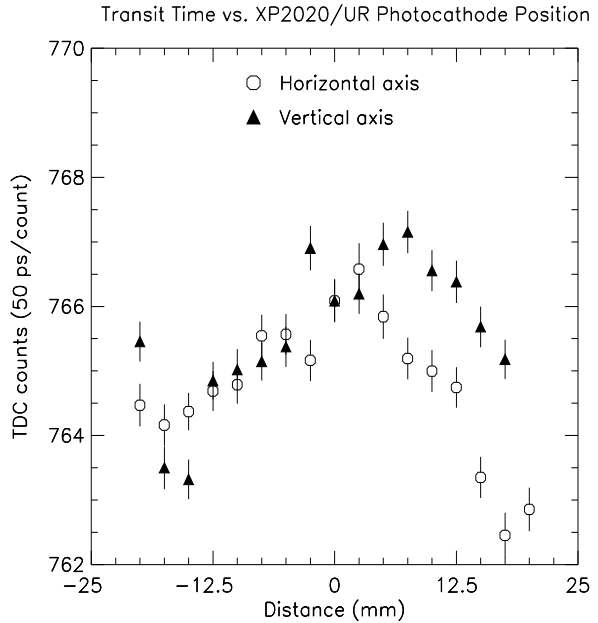,height=3.2in,width=3.2in}}
 \caption{Transit time dependence on the location of 
 the incident photon on the photocathode position for XP2020/UR.}
 \label{fig:082395_02}
\end{figure}

\begin{figure}
 \centering
 \mbox{\psfig{figure=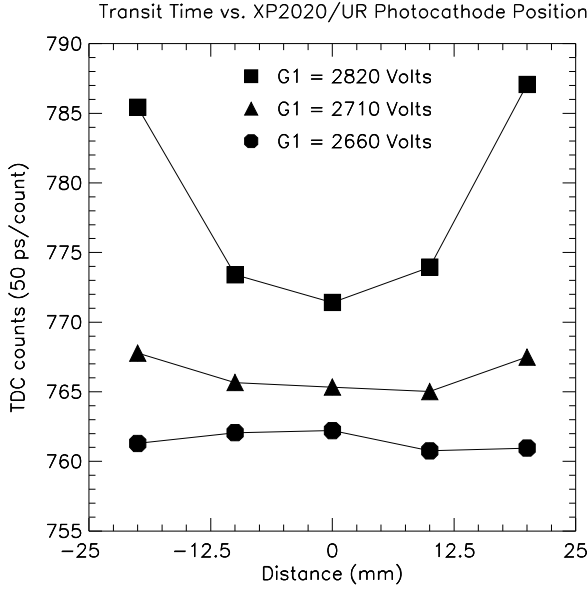,height=3.2in,width=3.2in}}
 \caption{Transit time vs the location of incident photon on 
  photocathode for  
 XP2020/UR at $-3000$V. Plotted for three 
 different potentials of the accelerating grid G1.}
 \label{fig:083195_02}
\end{figure}

Figure~\ref{fig:082395_02} shows the mean TDC stop time 
for horizontal and vertical scans of the photocathode.  
We have observed that the 
shape of this curve depends on the potential applied to the accelerating 
grid G1.  Figure~\ref{fig:083195_02} shows the mean corrected TDC stop time
as a function of horizontal beam location for three different G1 potentials.  
The position dependence of the transit time 
is clearly minimized for large potential differences between photocathode and
G1.  The variation of the shape with G1 potential suggests that the potential 
can be adjusted to provide a nearly uniform transit time for all 
photocathode positions, and that the resolution can be degraded if this 
parameter is not optimized.

The gain of each PMT/base combination was determined from the mean number of
ADC counts measured per pulse, accounting for the 0.25 pC/count calibration 
of the ADC and for the splitter.  The gain was measured at two different 
HV values, and found to vary as (HV)$^n$ with $n$ between 8.5 and 9.7.  This 
is consistent with the dependence expected for a twelve stage PMT.

\section{Monte carlo simulation}

We developed a Monte Carlo program to simulate the performance of the 
system with the XP2020/UR PMT's on the bulk counter in order to 
understand the relative importance of the physical processes contributing 
to the resolution of the system. 

Cosmic ray tracks were generated with a $\cos^2 (\alpha \theta)$ angular 
distribution where $\theta$ is the polar angle with respect to the
vertical.
The coefficient $\alpha = 2$ was chosen to match the shapes of
the reference counter pulse height as described later.
The number of photons was generated by the Landau distribution.  
The photon emission sites were 
distributed randomly along the length of the track, and the emission directions
were isotropic. The decay time of the scintillator was taken to be 2.1 ns.

The path of each photon was traced through the scintillator until the photon
either reached the PMT, or until it was lost by one of a 
number of different mechanisms.  Photons were absorbed in the scintillator 
assuming a bulk attenuation length of 3.8 meters for 
BC-408.  The index of refraction of BC-408 was taken to be 1.57 and the
reflection/transmission at boundaries were modeled
according to Fresnel's law for unpolarized light.
Furthermore, the photon was assumed to be lost at the boundary with
probability $1-R$ due to imperfection.
The photomultiplier window was modeled as a circular piece of 
glass with index of refraction 1.47 in direct contact with the scintillator.
The quantum efficiency of the photocathode was assumed to be 26\% 
averaged over the photon emission spectrum.  

\begin{figure}
 \centering
 \mbox{\psfig{figure=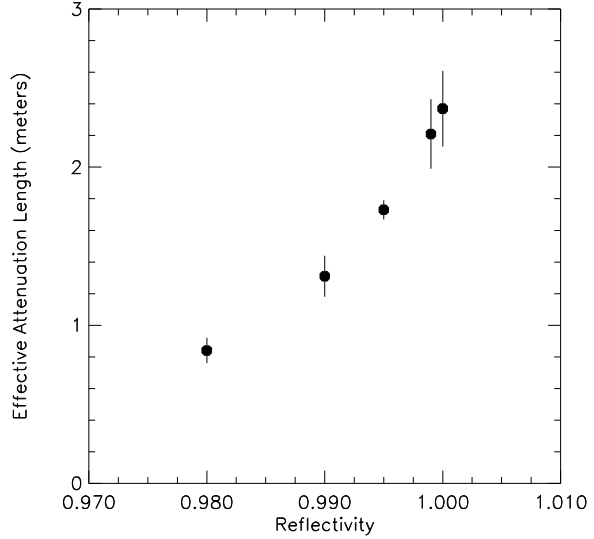,height=2.8in,width=3.2in}}
 \caption{Monte Carlo effective attenuation length of bulk counter as a 
  function of the reflectivity.  The best estimate of reflectivity was
  taken to be 0.9925.}
 \label{fig:042896_01}
\end{figure}

The reflectivity $R$ was chosen as follows:  
The effective attenuation length of the bulk counter 
is a function of the reflectivity, the bulk attenuation length, and the 
geometry of the counter. Figure~\ref{fig:042896_01} 
shows the effective attenuation length as a function of the reflectivity. 
We chose $R=0.9925$, which is the value that results in the measured effective 
attenuation length of 1.58 meters. 

Two transit time effects were modeled.  First, the transit time difference 
between center and edge of the photocathode was assumed to vary as $r^2$.  
For the reference counter, we took this difference to be 250 ps. 
For the XP2020/UR, we took the results from Figure~\ref{fig:082395_02}. 
Second, the transit time spread was modeled by 
a gaussian of sigma 250 ps.

The distribution of photoelectron arrival times was convoluted with
a function which described the time response of the PMT to a delta 
function light input.  For the test counter PMT's, this function was the 
pulse shape produced by the XP2020/UR on the single photon apparatus, 
measured by a 4 GHz Tektronix SCD5000 transient digitizer.  For the 
reference counters, this function was a gaussian of width 1.5 ns.  

The absolute yield of photons per unit path length was 
determined by 
requiring that the Monte Carlo pulse height distribution matches with
the data distributions. We find a 
production rate of 5890 photons/cm. This corresponds to a BC-408 light 
output which is 20\% of anthracene.
Using the resulting pulse shape for each track crossing, 
the discriminator triggering time was defined to be the time
the pulse shape crosses the threshold.

\begin{figure}
 \centering
 \mbox{\psfig{figure=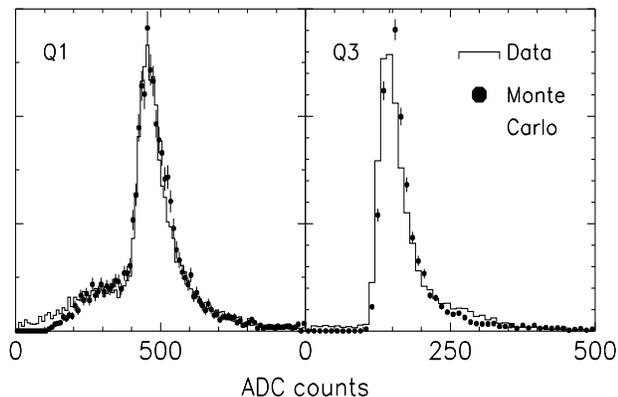,height=2.1in,width=3.2in}}
 \caption{Comparison of Monte Carlo and data for the pulse height
distributions of the reference counter (Q1) and the test counter (Q3).
The horizontal scales are adjusted by roughly 20\%\ to compare the
shapes.}
 \label{fig:MCDATAADC}
\end{figure}

The pulse height distributions of the reference counter (Q1) 
and the test counter (Q3) are shown for Monte Carlo and data in 
Figure~\ref{fig:MCDATAADC}.   
The lower shoulder for the reference counter
corresponds to tracks with large dip angle $\theta$ which intersect 
the sides of the reference counters. The coefficient 
$\alpha$ in the angular distribution $\cos^2 (\alpha \theta)$ was
determined to match the shape and size of this shoulder.
\begin{table}
\begin{center}
\begin{tabular}{llr}
\hline
(a) $\#\gamma$/cm   & \% of       & $\sigma_{\rm tof}$  \\
                    & anthracene  &      (ps)           \\
\hline
1440           & 5                 & $93 \pm 6$ \\
2880           & 10                & $72 \pm 6$ \\
5880 (nominal) & 20                & $55 \pm 3$ \\
8650	       & 29                & $37 \pm 7$ \\
11540	       & 38                & $32 \pm 9$ \\
23080	       & 77                & $33 \pm 5$ \\
\hline
\hline
(b) TTS (ps)   &    & $\sigma_{\rm tof}$ (ps) \\
\hline
1000          & & $91 \pm 3$ \\
 500          & & $61 \pm 4$ \\
 250 (nominal)& & $55 \pm 3$ \\
 125          & & $53 \pm 5$ \\
   0          & & $52 \pm 5$ \\
\hline
\hline
(c) Scintillation     & & $\sigma_{\rm tof}$ \\
    decay time (ps)   & &    (ps)    \\
\hline
 2.5                   &      & $55 \pm 6$ \\
 2.1 (nominal)         &      & $55 \pm 3$ \\
 1.75                  &      & $51 \pm 6$ \\
 1.5                   &      & $48 \pm 6$ \\
 1.0                   &      & $46 \pm 7$ \\
 0.0                   &      & $40 \pm 7$ \\
\hline
\hline
(d) Center/Edge      &   & $\sigma_{\rm tof}$ \\
    difference (ps)  &   &       (ps)         \\
\hline
400           &  & $56 \pm 4$ \\
300           &  & $55 \pm 4$ \\
200 (nominal) &  & $55 \pm 3$ \\
100           &  & $57 \pm 4$ \\
  0           &  & $55 \pm 4$ \\
\hline
\end{tabular}
\caption{Monte Carlo study:  Variation of the 
parameters in the simulation and the resulting change in
$\sigma_{\rm tof} (z = 100)$. The parameters varied are: (a)
number of photons produced per 
unit path length of track also expressed as
a percentage of anthracene output,
(b) the XP2020/UR transit time spread, 
(c) the BC-408 decay time,
(d) the XP2020/UR photocathode 
center/edge transit time difference.}
\end{center}
\label{tab:vary}
\end{table}

We calibrate the simulated data in the same way as described 
previously.  We find $\sigma_{\rm tof} = 55 \pm 3$ ps for 
the $z = 100$ cm trolley position, in reasonable agreement with the data.   

We varied a number of parameters in the Monte Carlo and observed the
resulting changes in $\sigma_{\rm tof}(z = 100)$.  The parameters
included the number of photons produced per unit length of cosmic 
ray track, the scintillator decay time, the transit time spread, and
the photocathode center-edge difference of the transit time.  In all cases,
the variations were applied to the test counter and its PMT's, but not to 
the reference counters.  Tables~\ref{tab:vary} 
summarizes these results.  We find that the resolution is limited primarily by 
photon statistics.  In our application, 
the resolution is fairly insensitive to the transit time 
difference between photocathode center and edge. This is because only the 
central portion of the photocathode of XP2020/UR is used. In
applications that use the entire photocathode, it will be critical to
adjust the potential of G1 to make the transit time uniform across the
photocathode surface. 

\section{Conclusions}

We have measured the resolution of a small cross section prototype 
time-of-flight system.  We find that a bulk counter manufactured to 
give an effective attenuation length comparable to a fiber counter 
gives a better resolution than the fiber counter due to the improved 
photon statistics.  Results using magnetic-field-resistant photomultiplier 
tubes are comparable to those obtained with a fast photomultiplier of a 
more conventional design.  We have studied the single photon response 
of XP2020/UR photomultiplier 
and found that the grid voltage adjustment is critical in
flattening the distribution of transit time as a function of 
photon position on photocathode.  
This is particularly important if the entire photocathode is to be 
illuminated.  We developed a detailed monte carlo simulation program
which reasonably 
reproduces the data.  The monte carlo shows that the critical
element is the number of photoelectrons, emphasizing 
in particular the importance 
of the internal  reflectivity of the scintillator block. 

\vskip .2in
{\large\bf Acknowledgement}

The authors would like to thank Ted Liu, Fay Shen, and technical staff at
the high energy physics laboratory at Harvard Univeristy for the
assistances at the early stage of the project.
This work was supported by the department of energy grant
DE-FG02-91ER40654.


\begin{thebibliography}{99}
\bibitem{othereffects}
In a typical colliding beam experiment,
one must also account for 
the finite size of the beam. For the CLEO detector, it 
introduces in quadrature an additional 30 ps uncertainty in the 
measured flight times.
\bibitem{MCestimate}
This technique does not account for the pulseheight
correlation due to tracks with large dip angle which has short path lengths
for both reference counters. Such pulseheight correlation couples to
the timing correlation through the time-walk correction.
Monte carlo studies indicate that 
we are underestimating the reference counter resolution by approximately 
10\%.  To be conservative, we do not apply this correction.
\end{thebibliography}
\end{document}